\documentclass[journal,twocolumn,10pt]{IEEEtran}
\usepackage{textcomp}
\usepackage{stfloats}
\usepackage{url}
\usepackage{graphicx}
\usepackage{verbatim}
\usepackage{upgreek}
\usepackage{amssymb,amsmath}
\usepackage{color}
\usepackage{epstopdf}
\usepackage{bm}
\usepackage{cite}
\usepackage{array,color}
\usepackage{algorithm}
\usepackage{algpseudocode}
\usepackage{amsmath}
\usepackage{amsfonts}
\usepackage{amsthm}
\usepackage{graphics}
\usepackage{epsfig}
\usepackage{soul}
\usepackage{gensymb}
\soulregister\cite7
\soulregister\ref7
\usepackage{subfigure}
\usepackage{setspace}

\newtheorem{rmk}{Remark}
\newtheorem{prop}{Proposition}

\begin{document}

\title{Revealing the Trade-off in ISAC Systems: \\The KL Divergence Perspective}

\author{Zesong~Fei,~\IEEEmembership{Senior~Member,~IEEE}, Shuntian~Tang, Xinyi~Wang,~\IEEEmembership{Member,~IEEE}, Fanghao~Xia, Fan~Liu,~\IEEEmembership{Senior~Member,~IEEE}, J.~Andrew~Zhang,~\IEEEmembership{Senior~Member,~IEEE}

\thanks{Zesong Fei, Shuntian Tang, Xinyi Wang and Fanghao Xia are with the School of Information and Electronics, Beijing Institute of Technology, Beijing 100081, China. (E-mail: feizesong@bit.edu.cn, 1120201688@bit.edu.cn, wangxinyi@bit.edu.cn, xiafanghaoxfh@163.com)}
\thanks{Fan Liu is with the School of System Design and Intelligent Manufacturing, Southern University of Science and Technology, Shenzhen 518055, China. (E-mail: liuf6@sustech.edu.cn)}
\thanks{J. Andrew Zhang is with the School of Electrical and Data Engineering, University of Technology Sydney, NSW, Australia 2007 (E-mail: Andrew.Zhang@uts.edu.au).}

\thanks{Corresponding author: Xinyi Wang.}

}



\maketitle

\begin{abstract}
Integrated sensing and communication (ISAC) {is regarded} as a promising technique for 6G communication network. In this letter, we investigate the Pareto bound of the ISAC system in terms of a unified Kullback-Leibler (KL) divergence performance metric. We firstly {present} the relationship between KL divergence and explicit ISAC performance metric, i.e., demodulation error and probability of detection. Thereafter, we investigate the impact of constellation and beamforming design on the Pareto bound via deep learning and semi-definite relaxation (SDR) techniques. Simulation results show the trade-off between sensing and communication performance in terms of bit error rate (BER) and probability of detection under different parameter set-ups.
\end{abstract}


\section{Introduction}

\IEEEPARstart{I}{ntegrated} sensing and communication (ISAC) technique has been viewed as a key enabler for achieving the paradigm of ``Intelligent Internet of Everything'' for 6G networks. Through integrating sensing and communication functionalities into a unified hardware platform and radio waveform over the same frequency band, energy-, spectral-, and hardware-efficiencies can be significantly improved, which has attracted more and more researchers to investigate the ISAC system design.

As a integrated system simultaneously considering both sensing and communication functionalities, the trade-off between them are of vital importance. Note that both sensing and communication systems have their typical performance indicators, e.g., capacity, bit error rate (BER) for communications, and detection probability, mean square error (MSE) for sensing \cite{Zhang2021JSTSP}. Hence, in existed literatures, transmission strategies have been designed based on various performance metrics. For example, in \cite{Liu2018TWC}, a pioneering ISAC beamforming strategy was proposed by considering the user signal-to-interference-plus-noise ratio (SINR) and transmit beampattern as sensing and communication performance metrics, respectively. Although the transmit beampattern is able to show the radiation power level towards diverse angles, it cannot directly reflect the sensing performance at the receiver side \cite{Liu2021TSP}. Therefore, in \cite{Liu2021TSP} and \cite{Wang2022TCOM}, the authors investigated the fully digital and hybrid analog-digital ISAC beamforming design, respectively, by adopting the Cramer-Rao bound (CRB) for angle estimation as the sensing performance metric. It has been shown that employing CRB as the performance metric achieves better estimation performance compared with the transmit beampattern based design. On the other hand, from the perspective of communication, the authors in \cite{Wang2023IoT} and \cite{johnston2022JSAC} respectively employed sum-rate and BER as performance metrics for ISAC beamforming design. As a step further, the authors in \cite{Xiong2023TIT} revealed the fundamental trade-off between communication rate and sensing CRB in Gaussian channels from the perspective of information theory.

Although the aforementioned studies have well addressed the performance trade-off between sensing and communication functionalities in ISAC systems, the performance metrics for sensing and communications therein are generally different. There is a lack of unified performance metric suited for both functionalities. Despite that the SINR is a crucial metric for both functionalities, it is not intuitively connected with the sensing performance. Under this background, the authors in \cite{Mohammad2023TWC} proposed a unified performance framework based on Kullback-Leibler (KL) divergence, which is closely related to the error rate performance of communication users and detection performance of sensing targets. However, the authors only analysed the performance of two existed beamforming schemes, without designing a more suitable transmission scheme.

In this letter, we utilize the KL divergence (KLD) as the design criterion for ISAC systems. We first define the unified performance metric based on KLD. Subsequently, we transform the unified KLD metric under the single-antenna set-up into a tractable form and investigate the design of constellation set for ISAC systems via deep learning. With the fixed constellation set, we show that the Pareto bound of the KLD region can be achieved through balancing between the eigenvectors corresponding to sensing subspace and communication subspace. Furthermore, we illustrate the trade-off between the BER and probability of detection.

\section{KLD-based Unified Performance Metric for ISAC Systems}

\begin{figure}
 \centering
 \includegraphics[width=0.4\textwidth]{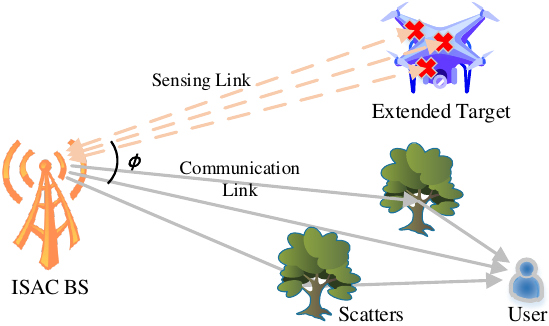} 
 \caption{An illustration of the considered ISAC system.}
 \label{system_model}
\end{figure}

The scenario of interest in this letter is illustrated in Fig. \ref{system_model}, where an $M$-antenna ISAC base station simultaneously serve a single-antenna user and perform sensing based on echoes for target detection. In particular, the Saleh-Valenzuela channel model, consisting of both one line-of-sight (LoS) and several non-LoS (NLoS) paths, is considered for the communication link, while the extended target model is considered for sensing.

\subsection{KLD in Communication Demodulation}

The signals received at the user can be expressed as
\begin{equation} \label{signal-c}
\begin{aligned}
y_c &= \sqrt{\rho_0 d_c^{-2}{P_t}} \left( \mathbf{a}^H(\theta_c) + \sum_{p=1}^{P}\alpha_p\mathbf{a}^H(\theta_p) \right)\mathbf{w}s + n_c \\
&\triangleq \sqrt{\rho_0 d_c^{-2}{P_t}} \mathbf{h}_c^H \mathbf{w}s + n_c,
\end{aligned}
\end{equation}
where $\rho_0$ denotes the path loss at the reference distance $d_0 = 1$ m, $d_c$ denotes the distance between the BS and communication user, {$P_t\mathbb{E}\{\vert s\vert^2\}$ denotes the average transmit power,} $\theta_c$ and $\theta_p$ denote the angles of departure (AoDs) of the line-of-sight (LoS) link and the $p$-th non-LoS (NLoS) link, $\mathbf{a}(\theta) = [1, e^{j2\pi \delta \sin(\theta)}, \cdots, e^{j2\pi(M-1)\delta \sin (\theta)}]^T$ is the antenna steering vector with $\delta$ denoting the normalized antenna spacing, $\alpha_p$ denotes the small-scale fading of the $p$-th path, $\mathbf{w}$ {with $\mathbf{w}^H\mathbf{w}=1$} denotes the normalized beamforming vector, $n_c \sim \mathcal{CN}(0, \sigma_c^2)$ denotes the additive white Gaussian noise (AWGN) at the user {red}{and $s$ denotes the transmitted symbol from the ISAC constellation set $\{c_1, c_2,..., c_Q\}$}.

Originally, the Kullback-Leibler Divergence (KLD) is defined for a pair of probability density functions (PDFs). Nevertheless, it can be extended multiple PDFs by considering every pair separately and evaluating the average {or the minimum} for all possible pairs. To be specific, for a generalized $Q$-ary signal constellation, KLD can be evaluated for each possible pair of unequal data symbols $\{c_m, c_n\}, m \neq n$ \cite{Mohammad2023TWC}. Note that the demodulation error performance is largely determined by the closest pair of symbols. Thus, the KLD in communication demodulation can be expressed as
\begin{align} \label{KLD-C}
\text{KLD}_c &= \min_{n \neq m} \text{KLD}_{m \rightarrow n} \\
&= \min_{n \neq m} \int_{-\infty}^{\infty} f_m(x) \log_2 \left( \frac{f_m(x)}{f_n(x)} \right) dx. \notag
\end{align}

Based on (\ref{signal-c}), the conditional density function of $y_c \vert \{c_m, \mathbf{w}\}$ can be expressed as
\begin{equation} \label{pdf}
f_m \triangleq f(y_c \vert \{c_m, \mathbf{w}\}) = \frac{\exp\left(-(\mathbf{y}_c - \bm{\mu}_m)^T \Sigma^{-1} (\mathbf{y}_c - \bm{\mu}_m)\right)}{\sqrt{(2\pi)^2\vert {\Sigma} \vert}},
\end{equation}
where $\mathbf{y}_c \triangleq [y_{c,\mathcal{R}}, y_{c, \mathcal{I}}]^T$ and $\bm{\mu}_m \triangleq [\mu_{m,\mathcal{R}}, \mu_{m,\mathcal{I}}]$, with $y_{c,\mathcal{R}} = \text{Re}\{y_c\}$, $y_{c,\mathcal{I}} = \text{Im}\{y_c\}$, $\mu_{m, \mathcal{R}} = \sqrt{\rho_0d_c^{-2}{P_t}}\text{Re}\{\mathbf{a}^H(\theta_c)\mathbf{w}c_m\}$ and $\mu_{m, \mathcal{I}} = \sqrt{\rho_0 d_c^{-2}{P_t}}\text{Im}\{\mathbf{a}^H(\theta_c)\mathbf{w}c_m\}${, $\Sigma= \sigma_c^2 \mathbf{I}_2$}.

By substituting (\ref{pdf}) into (\ref{KLD-C}), the KLD in communication demodulation can be derived as
\begin{align} \label{KLD_C}
\text{KLD}_c &= \min_{n \neq m} \frac{1}{2 \ln 2} \left( \text{Tr}(\Sigma_m^{-1} \Sigma_n) - 2  \right. \\
& \qquad\left. +(\bm{\mu}_m - \bm{\mu}_n)^H\Sigma_m^{-1}(\bm{\mu}_m - \bm{\mu}_n) +\ln \frac{\vert \Sigma_m \vert}{\vert \Sigma_n \vert} \right). \notag\\
& \overset{(a)}{=} \min_{n \neq m} \frac{{P_t}}{2\sigma_c^2 \ln 2}(\bm{\mu}_m - \bm{\mu}_n)^H(\bm{\mu}_m - \bm{\mu}_n), \notag
\end{align}
where $\overset{(a)}{=}$ holds since $\Sigma_n = \Sigma_m = \sigma_c^2 \mathbf{I}_2$.

\subsection{KLD in Radar Detection}

As for radar sensing, the detection of the target can be expressed as the following binary hypothesis testing problem:
\begin{equation} \label{received}
\mathbf{y}_r = \left\{ 
\begin{array}{ll}
\mathbf{n}_r, & \mathcal{H}_0, \\
\sqrt{\rho_0 d_t^{-4}{P_t}} \sum\limits_{j=1}^{J}{\mathbf{b}^{*}(\theta_j)}\mathbf{a}^H(\theta_j)\mathbf{w}s + \mathbf{n}_r, & \mathcal{H}_1,
\end{array} \right.
\end{equation}
where $d_t$ denotes the distance between the BS and target, {$\mathbf{b}(\theta)=\mathbf{a}(\theta)$ denotes the receive antenna steering vector,} $\theta_j$ denotes the AoD corresponding to the target, and $\mathbf{n}_r \sim \mathcal{CN}(\mathbf{0}, {{\sigma}_r^2 \mathbf{I}_M})$ denotes the AWGN at the BS. The PDF of $\mathbf{y}_r$ under $\mathcal{H}_0$ and $\mathcal{H}_1$ are
\begin{subequations}
\begin{align}
& f_0(\mathbf{y}_r) = \frac{1}{\pi^M \det (\mathbf{\Sigma}_n)} \exp(-\mathbf{y}_r^H \bm{\Sigma}_n^{-1} \mathbf{y}_r), \\
& f_1(\mathbf{y}_r) = \frac{1}{\pi^M \det (\mathbf{\Sigma}_n + \mathbf{\Sigma}_s)} \exp \left(-\mathbf{y}_r^H (\bm{\Sigma}_n + \mathbf{\Sigma}_s)^{-1} \mathbf{y}_r \right),
\end{align}
\end{subequations}
respectively, where $\mathbf{\Sigma}_s = \rho_0 d_t^{-4} {P_t}\mathbb{E}\{\vert s\vert^2\} \mathbf{A}\mathbf{W}\mathbf{A}^H$ with $\mathbb{E}\{\vert s\vert^2\} = \sum_{m=1}^Q \text{Pr}(c_m)\vert c_m \vert^2$, {$\text{Pr}(\cdot)$ denotes the probability, }$\mathbf{A} = \sum_{j=1}^J{\mathbf{b}^*(\theta_j)} \mathbf{a}^H(\theta_j)$ and $\mathbf{W} = \mathbf{w}\mathbf{w}^H$.

The KLD between $f_0(\mathbf{y}_r)$ and $f_1(\mathbf{y}_r)$ can then be derived as \cite{Tian2021TVT}
\begin{equation} \label{KLD_R}
\begin{aligned}
\text{KLD}_r &= \mathbb{E}\left\{ \ln \left( \frac{\det(\bm{\Sigma}_s + \bm{\Sigma}_n)}{\det(\bm{\Sigma_n})} \right. \right.\\
& \qquad \left. \times \exp\left( -\mathbf{y}_r^H(\bm{\Sigma}_n^{-1} - (\bm{\Sigma}_n+\bm{\Sigma_s})^{-1}) \right) \mathbf{y}_r \Big) \right\} \\
&= \ln \det\left( \mathbf{I}_M + \bm{\Sigma}_n^{-1/2}\bm{\Sigma}_s \bm{\Sigma}_n^{-1/2} \right) \\
& \quad + \text{Tr}\left( \left(\mathbf{I}_M + \bm{\Sigma}_n^{-1/2}\bm{\Sigma}_s \bm{\Sigma}_n^{-1/2} \right)^{-1} - \mathbf{I}_M \right).
\end{aligned}
\end{equation}

\subsection{Unified KLD-based Performance Metric}

Based on the expression in (\ref{KLD_C}) and (\ref{KLD_R}), the unified KLD-based performance metric for ISAC systems can be derived as
\begin{equation} \label{KLD_I}
\text{KLD}_I = (1-\eta) \text{KLD}_c + \eta \text{KLD}_r,
\end{equation}
where the weighting factor $\eta$ is used to balance between the communication performance and sensing performance.

Based on (\ref{KLD_I}), in what follows, we investigate the trade-off between the sensing and communication functionalities in terms of KLD. Considering that the maximum instantaneous transmission power is generally constrained in practice {to ensure the transmit power is within the linear dynamic range of amplifier}, we investigate the trade-off from the perspective of constellation and beamforming design under separate maximum power constraint.

\begin{rmk}
Note that although the unified KLD is related to both constellation set $\{c_m\}, m=1, \cdots, Q$ and beamforming vector $\mathbf{w}$, the constellation set for a fixed modulation order is typically fixed during transmission and required to be shared between the transmitter and receiver; otherwise the signalling overhead for sharing adopted constellation set will be inevitably high. Therefore, in what follows, we first propose the constellation design under the single-antenna set-up, and then investigate the beamforming design based on the optimized constellation set.
\end{rmk}

\section{Trade-off in Constellation Design under Single-antenna Set-up}

In this section, we investigate the constellation design under single-antenna set-up based on the unified KLD performance.

\begin{prop}
Under the single-antenna set-up, the unified KLD for ISAC systems can be simplified as
\begin{align}
	{\text{KLD}_I} = &\min_{n \neq m}\frac{(1-\eta)\rho_0 d_c^{-2} P_t}{2\sigma_c^2 \ln 2} \vert c_m - c_n \vert^2 \\
	& + \eta\left(\ln \left(1+\frac{\mathbb{E}\{ \vert c_m \vert^2\}\zeta}{\sigma_r^2} \right) + \frac{\sigma_r^2}{\mathbb{E}\{ \vert c_m \vert^2\}\zeta+\sigma_r^2} - 1\right), \notag
\end{align}
where $\zeta = \rho_0 d_t^{-4}P_t$.
\end{prop}

\begin{IEEEproof}
By denoting $c_m = \vert c_m \vert e^{j\psi_m}, m= 1, \cdots, Q$, the KLD for communication demodulation in (\ref{KLD_C}) can be simplified as

\begin{small}\begin{align}
\text{KLD}_c &= \min_{n \neq m}\frac{P_t}{2\sigma_c^2 \ln 2}(\bm{\mu}_m - \bm{\mu}_n)^H (\bm{\mu}_m - \bm{\mu}_n) \notag \\
&= \min_{n \neq m} \frac{\rho_0 d_c^{-2} P_t}{2\sigma_c^2 \ln 2} \big(\vert c_m \vert^2 + \vert c_n \vert^2 - 2\vert c_m \vert \vert c_n \vert\cos(\psi_m - \psi_n)\big) \notag \\
&= \min_{n \neq m} \frac{\rho_0 d_c^{-2} P_t}{2\sigma_c^2 \ln 2} \vert c_m - c_n \vert^2,
\end{align}\end{small}

As for radar sensing, the binary hypothesis for target detection can be simplified as
\begin{equation}
y_r = \left\{ 
\begin{array}{ll}
n_r, & \mathcal{H}_0, \\
\sqrt{\rho_0 d_t^{-4}{P_t}}s + n_r, & \mathcal{H}_1,
\end{array} \right.
\end{equation}
and the likelihood functions of $y_r$ under $\mathcal{H}_0$ and $\mathcal{H}_1$ is
\begin{subequations}
\begin{align}
& \tilde{f}_0(y_r) = \frac{1}{\pi \lambda_0} \exp\left( -\frac{\vert y_r\vert^2}{\lambda_0} \right), \\
& \tilde{f}_1(y_r) = \frac{1}{\pi \lambda_1} \exp\left( -\frac{\vert y_r\vert^2}{\lambda_1} \right),
\end{align}
\end{subequations}
respectively, where $\lambda_0 = \sigma_r^2$, and $\lambda_1 = \rho_0 d_t^{-4}P_t\mathbb{E}\{ \vert s \vert^2 \} + \sigma_r^2$. According to \cite{wang2023sensing}, the KLD between $\tilde{f}_0(y_r)$ and $\tilde{f}_1(y_r)$ is
\begin{equation} \label{KLD_r}
\begin{aligned}
\text{KLD}_r &= \ln \frac{\lambda_1}{\lambda_0} + \frac{\lambda_0}{\lambda_1} - 1 \\
&= \ln \left(1+\frac{\mathbb{E}\{ \vert c_m \vert^2\}\zeta}{\sigma_r^2} \right) + \frac{\sigma_r^2}{\mathbb{E}\{ \vert c_m \vert^2\}\zeta+\sigma_r^2} - 1.
\end{aligned}
\end{equation}
\end{IEEEproof}

Based on the unified KLD derived in \textbf{Proposition 1} {while ensuring the amplitude of transmitted symbol is within the linear dynamic range of amplifier}, the constellation design problem can be formulated as
\begin{align} \label{prob1}
\max_{\{c_m\}} \quad & \text{KLD}_I \\
\text{s.t.} \quad 
& \vert c_m \vert^2 \leq 1, m=1, \cdots, Q. \tag{\ref{prob1}a}
\end{align}

The problem (\ref{prob1}) is highly non-convex due to the complicated objective function. Nevertheless, as can be readily seen from (\ref{KLD_r}), the KLD for radar sensing is positively related to $\mathbb{E}\{\vert c_m \vert^2\}$ {as $f'(x)=1/(1+x)-1/(1+x)^2$ is always greater than $0$ when $x>0$.} As a result, $\text{KLD}_r$ can be maximized by maximizing $\mathbb{E}\{\vert c_m \vert^2\}$. Hence, a new and efficient unified KLD performance can be defined as
\begin{equation} \label{KLD_new}
\text{KLD}_{new} = (1-\eta_1) \min_{n \neq m} \vert c_m - c_n \vert^2 + \frac{\eta_1}{Q}\sum_{m=1}^Q \vert c_m \vert^2.
\end{equation}
Although $\text{KLD}_{new}$ is not equal to $\text{KLD}_I$, it can also achieve balance between sensing and communication performance by tuning $\eta_1$.

\begin{rmk}
The first term in (\ref{KLD_new}) measures inter-constellation distance, which determines the demodulation error, while the second term measures the average power of all constellation points, which determines the difference between the echo signals and noise while performing target detection based on energy detection. Therefore, the sensing and communication performance can be balanced through tuning $\eta_1$. Although the original $\text{KLD}_I$ is non-trivial to deal with, the value of factors $\frac{\rho_0 d_c^{-2} P_t}{2\sigma_c^2 \ln 2}$ and $\frac{\rho_0 d_t^{-4}P_t}{\sigma_r^2}$ can help determine the value of $\eta_1$ in practice.
\end{rmk}

\begin{figure}
 \centering
 \includegraphics[width=0.45\textwidth]{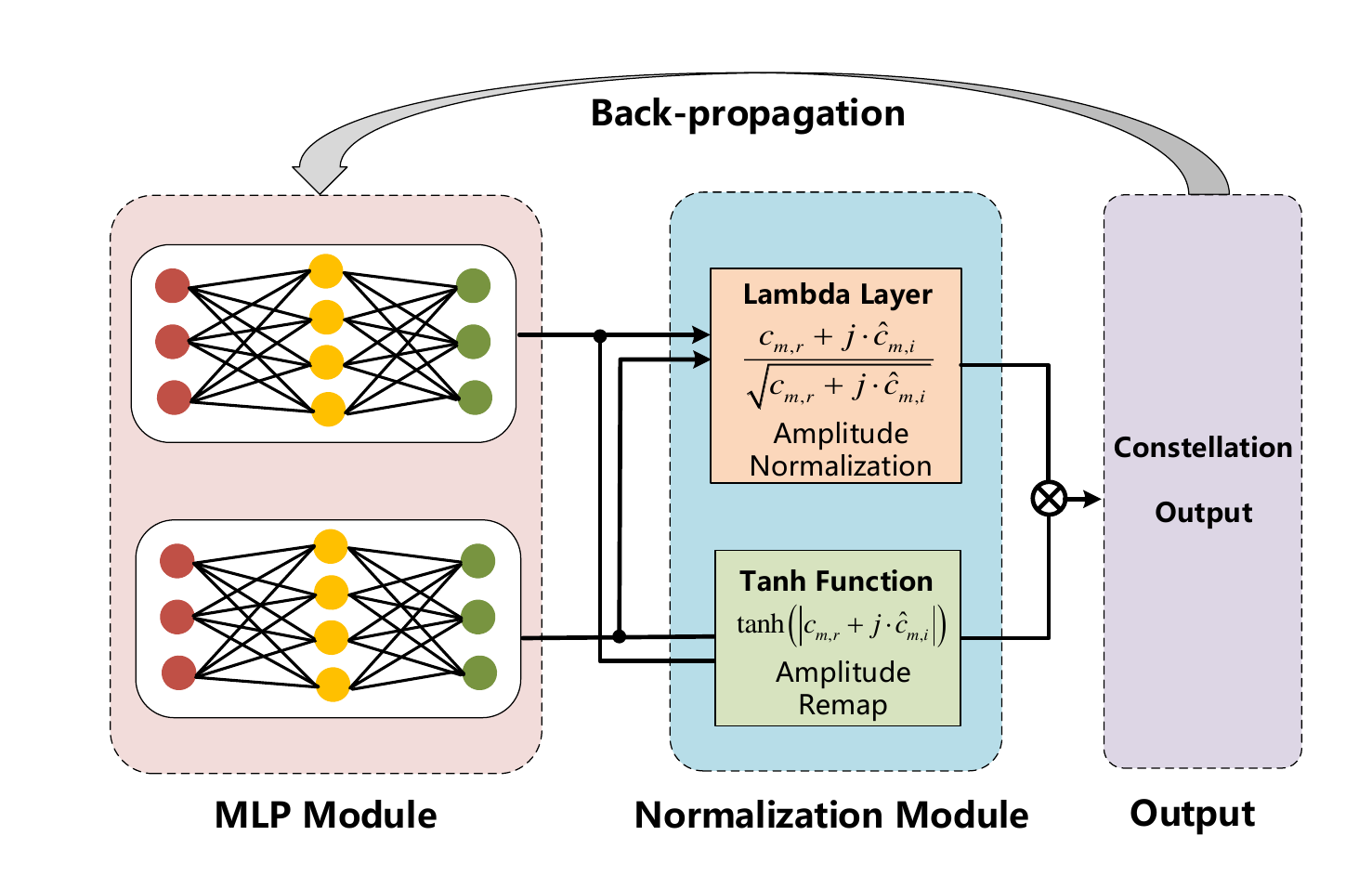} 
 \caption{Neural network architecture for constellation design.}
 \label{fig_neural_network}
\end{figure}

To address the non-convex constellation optimization problem, we propose a deep learning (DL)-based framework as shown in Fig. \ref{fig_neural_network}. The network consists of two multi-layer perceptron (MLP) modules and one normalization module. In particular, two MLP modules are respectively employed to generate the real part ${c}_{m,r}$ and imaginary part ${c}_{m,i}$ of each constellation point $c_m$, with each composed of five cascaded fully connected layers and using ReLU as the activation function. The output of two MLP modules are then combined in the normalizing layer as
\begin{align}
	\label{eq2.16}
	{c_m} = \tanh \left( {\left| {{c_{m,r}} + j \cdot {{\hat c}_{m,i}}} \right|} \right)\frac{{{c_{m,r}} + j \cdot {{\hat c}_{m,i}}}}{{\left| {{c_{m,r}} + j \cdot {{\hat c}_{m,i}}} \right|}},
\end{align}
where $\tanh(\cdot)$ denotes the tanh function to retract the amplitude into $[0,1]$. The unified KLD (\ref{KLD_new}) is adopted as the loss function for executing back propagation while training the network.

\section{Trade-off in KLD-based Beamforming Design}

Given the constellation set, the ISAC beamforming design {aims} to balance between sensing and communication functionalities. Therefore, in what follows, following the spirit in \cite{ouyang2023WCL}, where the impact of beamforming on the mutual information for sensing and communication was investigated, we first derive the optimal beamforming vectors for sensing and communication, and then investigate the ISAC design.

To further simplify $\text{KLD}_r$, we first introduce the Woodbury identity shown as follows
\begin{equation} 
(\mathbf{A}+\mathbf{UCV})^{-1}=\mathbf{A}^{-1}-\mathbf{A}^{-1}\mathbf{U}(\mathbf{C}^{-1}+\mathbf{V}\mathbf{A}^{-1}\mathbf{U})^{-1}\mathbf{V}\mathbf{A}^{-1},
\end{equation}
    where $\mathbf{A} \in \mathbb{C}^{M \times M}$, $\mathbf{U} \in \mathbb{C}^{M \times K}$, $\mathbf{C} \in \mathbb{C}^{K \times K}$, $\mathbf{V} \in \mathbb{C}^{K \times M}$. By replacing A and C with identity matrices, we can obtain
    \begin{equation} 
    (\mathbf{I}+\mathbf{UV})^{-1}=\mathbf{I}-\mathbf{U}(\mathbf{I}+\mathbf{VU})^{-1} \mathbf{V},
    \end{equation}
    let $\mathbf{U}=\mathbf{Aw}$, $\mathbf{V}=\beta\mathbf{w}^H\mathbf{A}^H$ and take the trace of both sides, we can obtain
    \begin{equation} 
    \text{Tr}(\mathbf{I}+\mathbf{UV})^{-1}=\frac{1}{1+\mathbf{VU}}+M-1.
    \end{equation}

Based on the above result, by omitting the constant, $\text{KLD}_r$ can be reformulated as
\begin{equation}
\begin{aligned}
\text{KLD}_r & = \ln(1+ \beta \mathbf{w}^H \mathbf{A}^H \mathbf{A}\mathbf{w}) + \frac{1}{1+\beta \mathbf{w}^H \mathbf{A}^H \mathbf{A}\mathbf{w}},
\end{aligned}
\end{equation}
where $\beta = \mathbb{E}\{\vert s \vert^2\} {\rho_0 d_t^{-4}{P_t}}/{\sigma_r^2}$. Again, note that $f(x)=\ln(1+x)+1/(1+x)$ is monotonically increasing with the increase of $x$; hence, the maximization of $\text{KLD}_r$ can be transformed as
\begin{equation} \label{prob-ws}
\max_{\mathbf{w}} \quad  \mathbf{w}^H \mathbf{A}^H \mathbf{A}\mathbf{w} \qquad  \text{s.t.} \quad \mathbf{w}^H \mathbf{w} =1.
\end{equation}

As can be readily seen, the optimal $\mathbf{w}$ for the problem (\ref{prob-ws}) $\mathbf{w}_s^* = {\mathbf{v}_{\max}}$, with $\mathbf{v}_{\max}$ denoting the eigenvector of $\mathbf{A}^H\mathbf{A}$ corresponding to the largest eigenvalue.

Similarly, the maximization of $\text{KLD}_c$ is equivalent to the maximization of $\mathbf{w}^H \mathbf{h}_c \mathbf{h}_c^H \mathbf{w}$. As a result, the optimal $\mathbf{w}$ from the perspective of communication is $\mathbf{w}_c^* = {\mathbf{u}_{max}} $, where $\mathbf{u}_{max} = {\mathbf{h}_{c}}$.

Consequently, the Pareto boundary of the $\{\text{KLD}_c, \text{KLD}_r\}$ region can be derived by solving the following problem.
\begin{align} \label{Pareto}
\max_{\mathbf{w}} \quad & \text{KLD}_r \\
\text{s.t.} \quad 
& \text{KLD}_c \geq \eta_2 \text{KLD}_{c}^{\max}, \tag{\ref{Pareto}a} \\
& \mathbf{w}^H \mathbf{w} =1, \tag{\ref{Pareto}b}
\end{align}
where $\text{KLD}_{c}^{\max}$ denotes the maximum $\text{KLD}_c$ with $\mathbf{w}=\mathbf{w}_c^*$, and different $\{\text{KLD}_c, \text{KLD}_r\}$ tuples can be obtained through adjusting the weighting factor $\eta_2 \in [0,1]$. Furthermore, note that $\text{KLD}_t, t\in \{r,c\}$ is positively related to $\mathbf{w}^H\mathbf{D}_t \mathbf{w}$, with $\mathbf{D}_r = \mathbf{A}^H\mathbf{A}$ and $\mathbf{D}_c = \mathbf{h}_c \mathbf{h}_c^H$. Therefore, the problem (\ref{Pareto}) can be equivalently transformed to the following form.
\begin{align} \label{Pareto1}
\max_{\mathbf{w}} \quad & \mathbf{w}^H \mathbf{D}_r \mathbf{w} \\
\text{s.t.} \quad 
& \mathbf{w}^H \mathbf{D}_c \mathbf{w} \geq \eta_2 \mathbf{w}_c^{*H} \mathbf{D}_c \mathbf{w}_c^*, \tag{\ref{Pareto1}a} \\
& \mathbf{w}^H \mathbf{w} =1. \tag{\ref{Pareto1}b}
\end{align}

According to \cite{Huang2010TSP}, the problem (\ref{Pareto1}) can be optimally solved through semidefinite relaxation (SDR) technique.

\begin{rmk}
Through tuning the factor $\eta_2$, the Pareto bound of the ISAC system can be derived. {Nevertheless, it should be noted that the Pareto bound is distinctive as $\mathbf{u}_{max}$ and $\mathbf{v}_{max}$ characterise the greatest eigenvector of sensing channel and communication channel, respectively, denoted as sensing subspace and communication subspace in\cite{lu2022performance}.  More specifically, through defining a correlation coefficient as $r = \mathbf{u}_{max}^H \mathbf{v}_{max}$, it can be readily seen that with a stronger correlation between the sensing and communication subspaces, the Pareto bound can be effectively improved, which will be illustrated via numerical results}.
\end{rmk}

\section{Numerical Results}

In this section, we provide numerical results via Monte-Carlo simulation to show {the} sensing-communication performance trade-off. Unless otherwise stated, the maximum transmit power of ISAC BS is set as 30 dBm, $\rho_0 = -30$ dB, the noise power at the BS and user are respectively set as -100 dBm and -80 dBm. The BS-user distance and BS-target distance are set as $d_c = 800$ m and $d_t = 1000$ m, respectively. The angle of the target is set as $0^\circ$, and the velocity of the target is set as 39 m/s. The number of NLoS links is set as $P = 4$, and the number of scattering point on the target is set as $J = 10$.

\begin{figure}
 \centering
 \includegraphics[width=0.5\textwidth]{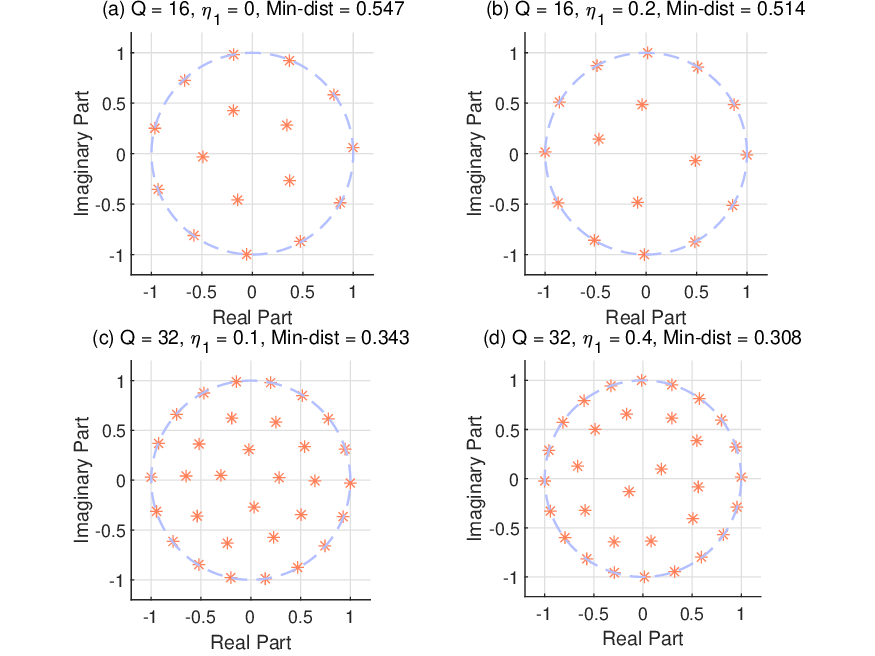} 
 \caption{Constellation under different modulation orders and weighting factors.}
 \label{fig_constellation}
\end{figure}

In Fig. \ref{fig_constellation}, we first show the optimized constellation set with different weighting factor $\eta_1$ in (\ref{KLD_new}). Two modulation orders are considered, i.e., $Q = 16$ and $Q = 32$. As can be seen, by designing the constellation set, we {actually conduct} geometric constellation shaping. Specifically, a smaller $\eta_1$ results in a larger minimum distance between adjacent constellations, corresponding to better symbol detection performance; while a larger $\eta_1$ results in higher average symbol power, corresponding to better target detection performance.

\begin{figure}
 \centering
 \includegraphics[width=0.35\textwidth]{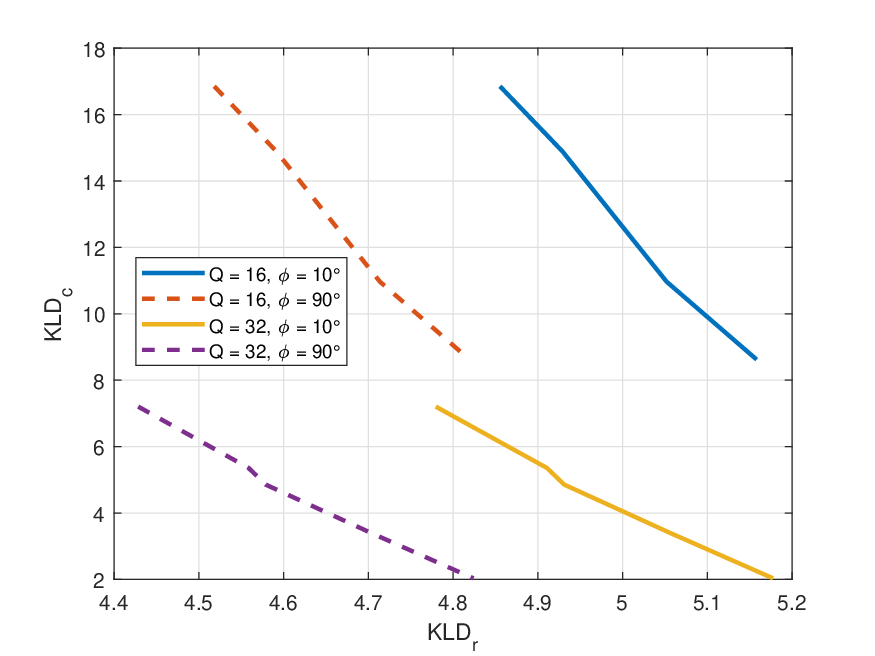} 
 \caption{Pareto bound between sensing and communication in terms of KLD.}
 \label{fig_KLD}
\end{figure}

To further reveal the performance trade-off, we illustrate the Pareto bound between sensing and communication performance in terms of KLD in Fig. \ref{fig_KLD}, with both constellation design and beamforming design taken into consideration. To show the diverse Pareto bound under different correlation between sensing and communication subspaces, two scenarios are considered. {In the first and second scenarios,} the angle between sensing link and communication link, i.e., $\phi$ in Fig. 1, is set as $10^\circ$ and $90^\circ$, corresponding to correlation coefficient $r_1 = 0.21$ and $r_2 = 0$, respectively. As can be seen, with a higher correlation coefficient, i.e., $\phi = 10^\circ$, a better Pareto bound can be achieved. Besides, the minimum distance with $Q = 16$ is larger than that with $Q = 32$, resulting in a larger $\text{KLD}_c$ with the same $\text{KLD}_r$. Nevertheless, it should be noted that {in the second scenario, a} higher throughput {can be achieved} under high SNR{s}. 

\begin{figure}
 \centering
 \includegraphics[width=0.35\textwidth]{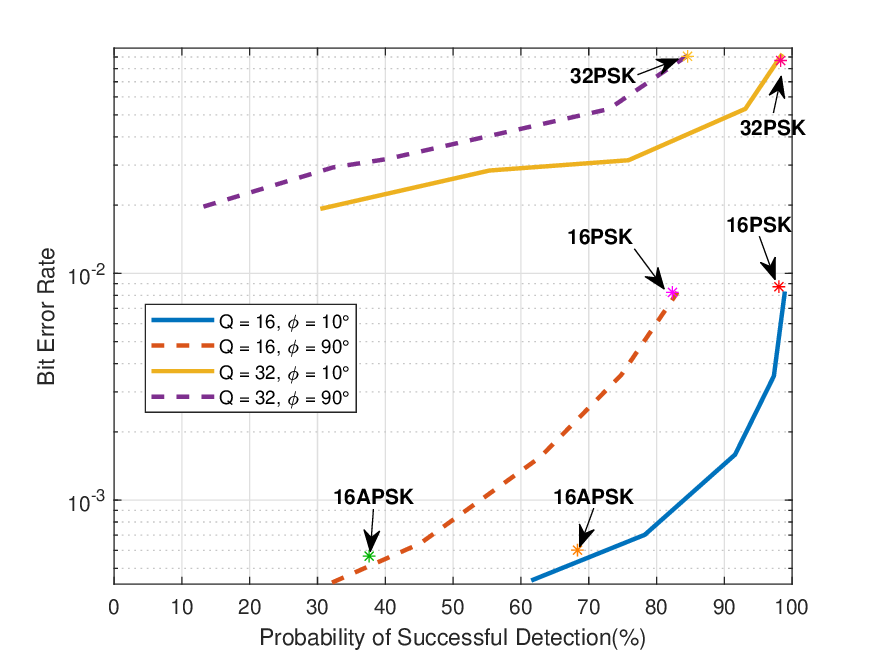} 
 \caption{Trade-off between BER and probability of detection.}
 \label{fig_P}
\end{figure}

As a step further, we depict the trade-off between BER and probability of detection in Fig. \ref{fig_P}. {In particular, two-dimensional constant false alarm rate (2D-CFAR) detection is executed with the false alarm probability set as $P_{\rm FA} = 10^{-5}$.} A total number of 16 pulses are accumulated to improve the signal-to-noise ratio (SNR) of the echo signals, with the length of each pulse being 40 $\micro$s. The same phenomenon as {in} Fig. \ref{fig_KLD} can be observed, and it can be seen that the difference{s} of the Pareto bound{s} under different $\phi$ {are} more significant with {a} higher modulation order. The reason is that, for a higher modulation order, the inter-symbol distance is smaller. In this case, the demodulation performance gain induced by beamforming with highly related sensing and communication channels is more significant. {To show the effectiveness of the Pareto bound, we also illustrate the sensing and communication capabilities of other commonly used constellations. It can be seen that the phase shift keying (PSK) modulation appears at one end of the bound, while the APSK modulation achieves {good} trade-off between sensing and communication.}

\section{Conclusion}

In this letter, we investigated the trade-off between sensing and communication functionalities for the ISAC system in terms of KL divergence. We first formulated the unified performance metric based on KL divergence, and then studied the design of constellation and beamforming via deep learning and the SDR technique. The Pareto bounds in terms of KL divergence as well as BER and probability of detection were finally provided to illustrate the trade-off between sensing and communication performance, under different correlation coefficients between sensing and communication links. {Numerical results show that with a stronger coupling between sensing and communication channels, a better Pareto bound can be achieved, which can provide insights for scheduling for ISAC systems with multiple targets and multiple communication users.}

\bibliographystyle{IEEEtran}%
\bibliography{bib/bibfile}

\vfill

\end{document}